\documentclass[10pt]{article}
\usepackage{amssymb}
\usepackage{graphicx}
\usepackage{layout}
\usepackage{latexsym}

\begin{document}

\title{Gravitomagnetic effects}
\author{M. L. Ruggiero, A. Tartaglia \\
Dip. Fisica, Politecnico and INFN, Torino, Italy\\
email: ruggierom@polito.it, tartaglia@polito.it}
\maketitle

\begin{abstract}
This paper contains a review of the theory and practice of gravitomagnetism,
with particular attention to the different and numerous proposals which have
been put forward to experimentally or observationally verify its effects.
The basics of the gravitoelectromagnetic form of the Einstein equations is
expounded. Then the Lense-Thirring and clock effects are described,
reviewing the essentials of the theory. Space based and Earth based
experiments are listed. Other effects, such as the coupling of
gravitomagnetism with spin, are described and orders of magnitude are
considered to give an idea of the feasibility of actual experiments.
\end{abstract}


\tableofcontents

\section{Introduction}

\label{sec:intro}

The close analogy between Newton's gravitation law and Coulomb's
law of electricity led many authors, in the past and also more
recently, to investigate further similarities, such as the
possibility that the motion of mass-charge could generate the
analogous of a magnetic field. The magnetic field is produced by
the motion of electric-charge, i.e. the electric current: the mass
current would produce what is called \textquotedblright
gravitomagnetic\textquotedblright\ field. Maxwell \cite{maxwell},
in one of his fundamental works on electromagnetism, turned his
attention to the possibility of formulating the theory of
gravitation in a form corresponding to the electromagnetic
equations. However, he was puzzled by the problem of the energy of
the gravitational field, i.e. the meaning and origin of the
negative energy due to the mutual attraction of material bodies.
In fact, according to him, the energy of a given field had to be
\textquotedblright essentially positive\textquotedblright , but
this is not the case of the gravitational field. To balance this
negative energy, a great amount of positive energy was required,
in the form of energy of the space. But, since he was unable to
understand how this could be, he did not proceed further along
this line.

On the ground of Weber's modification of the Coulomb law for the
electrical charges, Holzm\"{u}ller \cite{holz} first, then
Tisserand \cite{tiss}, proposed to modify Newton's law in a
similar way, introducing, in the radial component of the force law
a term depending on the relative velocity of the two attracting
particles (as it is well described by North \cite{north} and
Whittaker \cite{whittaker}). In 1893 Oliver Heaviside \cite{heavi1} \cite%
{heavi2} investigated the analogy between gravitation and
electromagnetism; in particular, he explained the propagation of
energy in a gravitational field, in terms of a
gravitoelectromagnetic Poynting vector, even though he (just as
Maxwell did) considered the nature of gravitational energy a
mystery. The formal analogy was then explored by Einstein
\cite{einstein1}, in the framework of General Relativity, and then
by Thirring \cite{thir1} who pointed out that the geodesic
equation may be written in terms of a Lorentz force, acted by a
gravitoelectric and gravitomagnetic field.

Any theory that combines Newtonian gravity together with Lorentz
invariance in a consistent way, must include a gravitomagnetic
field, which is generated by mass current. This is the case, of
course, of General Relativity: it was shown by Lense and Thirring
\cite{lensthir1}, that a rotating mass generates a gravitomagnetic
field, which in turn causes a precession of planetary orbits. It
is indeed interesting to notice that Lodge and Larmor, at the end
of the nineteenth century discussed the effects of frame dragging
on a non rotating interferometer \cite{ande}, but within the
framework of an aether-theoretic model This frame dragging
corresponds in fact to the Lense-Thirring effect of General
Relativity. However, at the beginning of the XX century, when
Lense and Thirring published their famous papers, the effect named
after them, which is indeed very small in the terrestrial
environment, was far from being detectable, because of the
technical difficulties and limitations of the time.

Contemporary improvements in technology have made possible to
propose new ideas to reveal the Lense-Thirring precession by
analyzing the data sets on the orbits of Earth satellites
\cite{ciufo}. An actual experiment to observe the effects of the
gravitomagnetic field of the Earth is about to fly; this is GP-B
\cite{gpb}, which is described further on in this paper. Other
experiments (like GP-C \cite{gpc}) have been proposed to reveal
the space-time structure, which is affected by gravitomagnetism,
for example evidencing clock effects around a spinning massive
object.\newline

The main purpose of this paper is to review the various old and
new tests of gravitomagnetism, which have been proposed in the
past and may be viable today, or in the next years. Considering
the growing of technological skills, many experiments which were
pure science-fiction ideas some decades ago, or nothing more than
\textit{gedanken experimente}, deserve today a
detailed study. They would both represent new tests of General Relativity%
\footnote{%
Of course, experiments could in principle discriminate between
General Relativity and other theories of gravitation.} and give us
the possibility of understanding how deep the analogy between the
electrical and gravitational phenomena is.

In Section \ref{sec:gem} we establish the theoretical apparatus,
starting from Einstein's equations in weak field approximation, to
obtain the gravitoelectromagnetic equations. Then we present the
tests of the rotation effects, starting from the Lense-Thirring
effect which we analyze in details in Section \ref{sec:lt}. In
Sections \ref{sec:clock} and \ref{sec:sagnac} we focus on the
effects of rotation on clocks. The effect on signals propagation
in both the terrestrial and the astrophysical environment are
presented in Sections \ref{sec:MM} and \ref{sec:signals}. The
possibility of detecting the coupling of intrinsic Spin with the
gravitomagnetic field is reviewed in Section \ref{sec:coupling}.
The raising of gravitomagnetic forces and their effects is
presented in Section \ref{sec:forces}, while other gravitomagnetic
effects are listed in Section \ref{sec:other}.

Finally, in Section \ref{sec:conclusion} we sketch the
conclusions, stressing the possibility that a number of the
effects presented in this paper could be revealed in the
foreseeable future, taking into account both the improvement in
technologies and the start of new promising research projects.

\section{From Einstein field equations to Gravitoelectromagnetism}

\label{sec:gem}

Einstein's equations may be written in a very simple way, which
leads straight to the analogy with Maxwell's equations, if we
consider the so called weak field approximation. We can use this
kind of linear approximation if we deal with a source whose
gravitational field is weak and, if the source is rotating, its
rotation is not relativistic. \newline With the partial exception
of the binary pulsar PSR 1913+16 (\cite{straumann} sec. 5.6), the
weak field is the normal condition for all the tests of General
Relativity up to this moment. \newline We shall follow the
standard treatment. So, let us start from the full non
linear equations\footnote{%
Notation: latin indices run from 1 to 3, while greek indices run
from 0 to 3; the flat space time metric tensor is $\eta _{\mu \nu
}=diag(1,-1,-1,-1)$,
in cartesian coordinates $x^{0}=ct$, $x^{1}=x$, $x^{2}=\allowbreak y$, $%
x^{3}=z$; boldface letters refer to space vectors, like the
position of a point $\mathbf{x}$.}.
\begin{equation}
G_{\mu \nu }=8\pi \frac{G}{c^{4}}T_{\mu \nu }  \label{eq:einstein}
\end{equation}
We put
\begin{equation}
g_{\mu \nu }=\eta _{\mu \nu }+h_{\mu \nu }  \label{eq:metric}
\end{equation}
where $\eta _{\mu \nu }$ is the Minkowski metric tensor, and
$|h_{\mu \nu }|\ll 1$ is a ''small''\ deviation from
it\footnote{$|h_{\mu \nu }|\ll 1$:
as we shall see, in the solar system $|h_{\mu \nu }|\simeq |\frac{\Phi }{%
c^{2}}|\leq 10^{-6}$, where $\Phi $ is the newtonian potential.
This makes
the approximation consistent.}. Then we define\footnote{%
In this linear approximation, space time indices are lowered and
raised using the flat-space time metric tensor $\eta _{\mu \nu }$;
in particular
space indices are raised and lowered by means of the euclidean tensor $%
\delta _{ij}=(1,1,1)$ .}
\begin{equation}
\overline{h}_{\mu \nu }=h_{\mu \nu }-\frac{1}{2}\eta _{\mu \nu
}h\mathit{;}\ \ h=h_{\ \alpha }^{\alpha }  \label{hbar}
\end{equation}
So if we expand the field equations (\ref{eq:einstein}) in powers of $%
\overline{h}_{\mu \nu }$ keeping only the linear terms, we obtain
(\cite{MTW} sec. 18.1)
\begin{equation}
\square \overline{h}_{\mu \nu }=16\pi \frac{G}{c^{4}}T_{\mu \nu }
\label{eq:einlinear}
\end{equation}
where we have also imposed the so called Lorentz gauge condition $\overline{h%
}_{\ \ \ ,\alpha }^{\ \mu \alpha }=0$. Equations
(\ref{eq:einlinear}) constitute the ''linearized''\ Einstein's
field equations. The analogy with the corresponding Maxwell
equations
\begin{equation}
\square A^{\nu }=4\pi j^{\nu }  \label{eq:max1}
\end{equation}
is evident.

The solution of (\ref{eq:einlinear}) may be written exactly in
terms of retarded potentials:
\begin{equation}
\overline{h}_{\mu \nu }=-4\frac{G}{c^{4}}\int \frac{T_{\mu \nu }(t-|\mathbf{x%
}-\mathbf{x^{\prime }}|/c,\mathbf{x^{\prime }})}{|\mathbf{x}-\mathbf{%
x^{\prime }}|}d^{3}x^{\prime }  \label{eq:potrit}
\end{equation}
The role of the electromagnetic vector potential $A^{\nu }$ is
played here by the tensor potential $\overline{h}_{\mu \nu }$,
while the role of the
four-current $j^{\nu }$ is played by the stress-energy tensor $T^{\mu \nu }$.%
\newline

We look for solutions such that $|\overline{h}_{00}|\gg |\overline{h}_{ij}|$%
, $|\overline{h}_{0i}|\gg |\overline{h}_{ij}|$ (Mashhoon
\textit{et. al} \cite{mashh1}), and neglect the other terms, which
are smaller.\newline The explicit expression for the tensor
potential $\overline{h}_{\mu \nu }$ is then:
\begin{eqnarray}
\overline{h}^{00} &=&\frac{4\Phi }{c^{2}}  \label{eq:potM} \\
\overline{h}^{0l} &=&-2\frac{A^{l}}{c^{2}}  \label{eq:potJ}
\end{eqnarray}
Where $\Phi $ is the Newtonian or ''gravitoelectric''\ potential
\begin{equation}
\Phi =-\frac{GM}{r}  \label{eq:potM2}
\end{equation}
while $\mathbf{A}$ is the ''gravitomagnetic''\ vector potential in
terms of the total angular momentum of the system
$\mathbf{S}$\footnote{$\varepsilon _{ijk} $ is the
three-dimentional completely antisymmetric tensor of
Levi-Civita.}:
\begin{equation}
A^{l}=\frac{G}{c}\frac{S^{n}x^{k}}{r^{3}}\varepsilon _{nk}^{l}
\label{eq:potJ3}
\end{equation}
It follows that $\frac{T^{00}}{c^{2}}=\rho $ is the
''mass-charge''\ density. Hence the total mass $M$ of the system
is
\begin{equation}
\int \rho d^{3}x~=~M  \label{eq:M}
\end{equation}
while $\frac{T^{i0}}{c}~=~j^{i}$ represents the mass-current
density, and the total angular momentum of the system is
\begin{equation}
S^{i}=2\int \varepsilon _{jk}^{i}x^{\prime }{}^{j}\frac{T^{k0}}{c}%
d^{3}x^{\prime }  \label{eq:potJ2}
\end{equation}
In terms of the potentials $\Phi ,\mathbf{A}$, the Lorentz gauge
condition becomes
\begin{equation}
\frac{1}{c}\frac{\partial \Phi }{\partial t}+\frac{1}{2}\mathbf{\nabla }%
\cdot \mathbf{A}=0  \label{eq:gaugeAA}
\end{equation}
which, apart from a factor $1/2$, is the Lorentz condition of
electromagnetism\footnote{%
In this equation and in the following ones, there is a factor
$1/2$ which does not appear in standard electrodynamics: the
effective gravitomagnetic charge is twice the gravitoelectric one;
this is a remnant of the fact that the linear approximation of GR
involves a spin-2 field, while ''classical''\
electrodynamics involves a spin-1 field (see Wald \cite{wald}, section 4.4)}.%
\newline
It is then straightforward to define the gravitoelectric and
gravitomagnetic fields $\mathbf{E}_{G}$, $\mathbf{B}_{G}$:\emph{\
}
\begin{eqnarray}
\mathbf{E}_{G} &=&-\mathbf{\nabla }\Phi -\frac{1}{2c}\frac{\partial \mathbf{A%
}}{\partial t}  \label{eq:EAA} \\
\mathbf{B}_{G} &=&\mathbf{\nabla }\wedge \mathbf{A} \label{eq:BAA}
\end{eqnarray}
Using equations (\ref{eq:einlinear}), (\ref{eq:gaugeAA}), (\ref{eq:EAA}), (%
\ref{eq:BAA}), and the definitions of mass density and current, we
finally get the complete set of Maxwell's equations for the so
called gravitoelectromagnetic (GEM) fields:\emph{\ }
\begin{eqnarray}
\mathbf{\nabla }\cdot \mathbf{E}_{G} &=&-4\pi G\rho  \label{eq:ME1} \\
\mathbf{\nabla }\cdot \mathbf{B}_{G} &=&0  \label{eq:MB2} \\
\mathbf{\nabla }\wedge \mathbf{E}_{G} &=&-\frac{1}{2c}\frac{\partial \mathbf{%
B}_{G}}{\partial t}  \label{eq:ME2} \\
\mathbf{\nabla }\wedge \frac{1}{2}\mathbf{B}_{G} &=&\frac{1}{c}\frac{%
\partial \mathbf{E}_{G}}{\partial t}-\frac{4\pi G}{c}\mathbf{j}
\label{eq:MB1} \\
\mathbf{E}_{G} &=&-\mathbf{\nabla }\Phi -\frac{1}{2c}\frac{\partial \mathbf{A%
}}{\partial t}  \label{eq:E} \\
\mathbf{B}_{G} &=&\mathbf{\nabla }\wedge \mathbf{A}  \label{eq:B} \\
\frac{1}{c}\frac{\partial \Phi }{\partial t}+\frac{1}{2}\mathbf{\nabla }%
\cdot \mathbf{A} &=&0  \label{eq:gauge}
\end{eqnarray}
Einstein's field equations in this form correspond to a solution
that describes the field around a rotating object in terms of
gravitoelectric and gravitomagnetic potentials; the metric tensor
can be read from the corresponding space-time invariant:\emph{\ }
\begin{equation}
ds^{2}=(1+2\frac{\Phi }{c^{2}})c^{2}dt^{2}+4dt(\mathbf{dr}\cdot \frac{%
\mathbf{A}}{c})-(1-2\frac{\Phi }{c^{2}})\delta _{ij}dx^{i}dx^{j}
\label{eq:metr}
\end{equation}
Hence the gravitational field is understood in analogy with
electromagnetism. For instance, the gravitomagnetic field of the
Earth as well as of any other weakly gravitating and rotating mass
may be written as a dipolar field:\emph{\ }
\begin{equation}
\mathbf{B}_{G}=-4\frac{G}{c}\frac{3\mathbf{r}(\mathbf{r}\cdot \mathbf{S})-%
\mathbf{S}r^{2}}{2r^{5}}  \label{eq:Bdip}
\end{equation}
\emph{\newline
\newline
}To complete the picture, a further analogy can be mentioned. In
fact, using the present formalism, the geodesic equation for a
particle in the field of a weakly gravitating, rotating object,
can be cast in the form of an equation of motion under the action
of a Lorentz Force.\newline The geodesic equation is\emph{\ }
\begin{equation}
\frac{d^{2}x^{\alpha }}{ds^{2}}+\Gamma _{\mu \beta }^{\alpha }\frac{dx^{\mu }%
}{ds}\frac{dx^{\beta }}{ds}=0  \label{eq:geo}
\end{equation}
If we consider a particle in non-relativistic motion, we have $\frac{dx^{o}}{%
ds}\simeq 1$, hence the velocity of the particle becomes $\frac{v^{i}}{c}%
\simeq \frac{dx^{i}}{ds}$; at the same time we neglect terms in the form $%
\frac{v^{i}v^{h}}{c^{2}}$. Limiting ourselves to static fields, where $%
g_{\alpha \beta ,0}=0$, it is easy to verify that the geodesic
equation may be written as\
\begin{equation}
\frac{d\mathbf{v}}{dt}=\mathbf{E}_{G}+\frac{\mathbf{v}}{c}\wedge \mathbf{B}%
_{G}  \label{eq:geolorentz}
\end{equation}
which shows that free fall, in the field of a massive rotating
object, can be looked at as motion under the action of the Lorentz
force produced by the GEM fields.\newline

This is the basic background in which all tests of GEM take place.
Even if we are more interested in the rotation effects, that is
the gravitomagnetic effects, in the following sections we shall
refer to gravitoelectromagnetic fields, which include also the
gravitoelectric, or newtonian, part.\newline It is useful to
exploit the analogy with electromagnetism, because, as we will see
in a while, it simplifies the solutions of some problems. On the
other hand, it is important to understand how far reaching this
correspondence is.

\section{The Lense-Thirring effect}

\label{sec:lt}

The most famous gravitomagnetic effect is indeed the Lense and
Thirring effect. Here we shall review both the effect per se, and
the existing and proposed tests of it.\\

\emph{The Mach Principle}.
\addcontentsline{toc}{subsubsection}{The Mach Principle} As we
said in the Introduction, at the beginning of the XX century,
Josef Lense and Hans Thirring \cite{lensthir1}, studied the
effects of rotating masses within the relativistic theory of
gravitation. Their
starting intention was to incorporate the so called ''Mach's principle''\cite%
{mach} into the GR theory. Mach's belief was that the origin of
inertia was in the global distribution of matter in the universe.
A consequence of this assumption was that moving matter should
somehow drag with itself nearby bodies. In particular rotating
matter should induce rotation and specifically cause the
precession of the axis of a gyroscope: a spinning body, which
creates a gravitomagnetic field, somehow drags the free frame,
which is a gyroscope.

Mach's conjecture is fascinating, though, up to now, a definite
and incontroversial incorporation of the principle into GR has not
been attained. A more accurate analysis can be found in
\cite{mashh1} and in the references therein.\newline

\emph{The effect}. \addcontentsline{toc}{subsubsection}{The
effect}The 'gyroscopes' that Lense and Thirring firstly took into
account were planets and natural satellites acted upon by the
rotating Sun, or Earth. The induced precession should show up in
the orbital angular momentum of planets and satellites. The
expected effect is indeed small, but, using artificial satellites
with appropriately chosen orbits, it could appear as a precession
of the orbital plane around the rotation axis of the Earth.

Let us start from the beginning. The metric tensor in the vicinity
of a spinning mass in weak field approximation, may be read off
from the space-time invariant (\cite{ruffi} ch. 4)
\begin{eqnarray}
ds^{2} &=&\left( 1-\frac{2GM}{rc^{2}}\right) c^{2}dt^{2}-\left( 1+\frac{2GM}{%
c^{2}r}\right) dr^{2}-r^{2}d\theta ^{2}-r^{2}\sin ^{2}\theta d\phi
^{2}
\label{eq:debole} \\
&&+\frac{4GMa}{cr}\sin ^{2}\theta d\phi dt  \nonumber
\end{eqnarray}
Here $a=S/Mc$, i.e. $a$, apart from a $c$ factor, represents the
angular momentum of the source per unit mass; more precisely
stated, it is the projection of the angular momentum three-vector
on the direction of the rotation axis, divided by the mass and the
speed of light.\newline To calculate the entity of the looked for
precession rate, we have to solve the equations of motion of a
test body in the gravitational field corresponding to the metric
tensor (\ref{eq:debole}): it is what Lense and Thirring did in
their papers. It is however simpler and clearer to use the GEM
equations (\cite{ruffi} ch.4).\newline

The Lense-Thirring precession is analogous to the precession of
the angular momentum of a charged particle, orbiting about a
magnetic dipole; the orbital momentum of the particle divided by 2
and by $c$ will then be the
equivalent of the magnetic dipole moment of the particle\footnote{%
Here the particle is assumed not to be itself a gyroscope, i.e. it
is spinless.}. So, if\emph{\normalsize \ }
\begin{equation}
\mathbf{B}_{G}=-4\frac{G}{c}\frac{3\mathbf{r}(\mathbf{r}\cdot \mathbf{S})-%
\mathbf{S}r^{2}}{2r^{5}}  \label{eq:Bdip1}
\end{equation}
is the gravitomagnetic field of the Earth ($\mathbf{S}$ is the
angular momentum of the Earth) the torque on the angular momentum
$\mathbf{L}$ of the orbiting particle is
\begin{equation}
\mathbf{\tau }=\frac{\mathbf{L}}{2c}\wedge \mathbf{B}_{G}
\label{eq:LT1}
\end{equation}
The time derivative of the orbital angular momentum can then be written as%
\emph{\normalsize \ }
\begin{equation}
\frac{d\mathbf{L}}{dt}=-G\frac{\mathbf{L}\wedge (3\mathbf{r}\mathbf{r\cdot S}%
-\mathbf{S}r^{2})}{c^{2}r^{5}}  \label{eq:LT2}
\end{equation}
From (\ref{eq:LT2}) we read the angular velocity of the precession ($\frac{d%
\mathbf{L}}{dt}=\mathbf{\Omega }\wedge \mathbf{L}$)
\begin{equation}
\mathbf{\Omega }=G\frac{3\mathbf{r}\mathbf{r\cdot S}-\mathbf{S}r^{2}}{%
c^{2}r^{5}}  \label{eq:LT3}
\end{equation}
If we take the average of $\mathbf{\Omega }$ along the orbit, the
effective angular velocity of precession is\emph{\normalsize \ }
\begin{equation}
<\mathbf{\Omega >}=G\frac{3<\mathbf{r}\mathbf{r\cdot S}>-\mathbf{S}r^{2}}{%
c^{2}r^{5}}  \label{eq:LT33}
\end{equation}
For an orbit with $r\simeq R_{Earth}$ the magnitude of the
precession rate is about $0.05\ $arcsec per year.\newline

\subsection{Observational verification}

\label{ssec:Obs}

\emph{Observational difficulties.} %
\addcontentsline{toc}{subsubsection}{Observational difficulties}
The gravitomagnetically induced orbital precession of artificial
satellites is smaller then the effects caused by another kind of
perturbation, such as the one generated by the quadrupole moment
of the Earth. To detect it, however, many witty proposals have
been made. For example Van Patten and Everitt proposed to use two
polar satellites to eliminate the effect of the Earth's quadrupole
\cite{vaneve}. Ciufolini proposed to use two non polar satellites,
with opposite inclinations, for the same purpose \cite{ciufo}.
\newline

\emph{Direct evidences.}
\addcontentsline{toc}{subsubsection}{Direct evidences}Analysing
the existing laser ranging observations of the orbits of
the satellites LAGEOS and LAGEOS II, Ciufolini et al. \cite{ciufo1},\cite%
{ciufo2} obtained the first direct \normalfont measurement of the
gravitomagnetic orbital perturbation due to the Earth's rotation,
i.e. the Lense-Thirring effect, within an accuracy of
20-30\%.\newline

\emph{A further proposal}. \addcontentsline{toc}{subsubsection}{A
further proposal} Further improvements in the obervation of the
angular
precession of the orbital plane could be attained with the LARES \cite%
{ciufo3} proposed mission, which would be able to test the
Lense-Thirring precession within an accuracy of 2-3\% and would
also be useful for a better understanding of fundamental
gravitational phenomena such as testing the inverse square law,
the equivalence principle, the PPN parameters etc. A new scenario
for the LARES mission has been recently described by Iorio \textit{et al.}\cite{iorio02}.%
\newline

\emph{Indirect evidences}.
\addcontentsline{toc}{subsubsection}{Indirect evidences} Indirect
evidences of the Lense-Thirring dragging of the inertial frames
or, more generally, of gravitomagnetic influences were given by
the periastron precession rate of the binary pulsar PSR 1913+16,
and also by
studying the laser ranging of the orbit of the Moon \cite{nordve},\cite%
{shahid},\cite{cliff}.\newline We consider these observations as
indirect because the sought for precession is in general combined
with other effects some of which are not immediately disentangled
from each other. Anyway the evidence for gravitomagnetic
perturbations, considered as post newtonian forces between moving
bodies, can nowadays be considered as clearly acquired,
notwhithstanding some contrary impressions still diffused a few
years ago, as K. Nordtvedt points out \cite{nordve1}.\newline

\emph{\normalsize Lense - Thirring and QPO }%
\addcontentsline{toc}{subsubsection}{Lense - Thirring and QPO} The
effects of the Lense - Thirring precession are of some interest in
the study of
quasi-periodic oscillations (QPO) in black holes \cite{merlo},\cite{comi},%
\cite{cui}.\newline

\subsection{Space based experiments}

\label{ssec:spacebased}

\emph{Gravity Probe B}.
\addcontentsline{toc}{subsubsection}{Gravity Probe B} Gravity
Probe B (GP-B) is an experiment whose purpose is to detect the
Lense-Thirring effect by measuring the precession rate of an
orbiting
gyroscope\footnote{%
Whereas in the other examples the whole of the satellite's orbit
was to be
considered as a gyroscope.}. It was proposed firstly by Schiff \cite{schiff}%
, and then developed by Fairbank, Everitt et al.
\cite{gpb},\cite{gpb2}. The actual launch of the satellite, which
will carry four gyroscopes, is scheduled for the nearest future
\cite{gpb3}.

The precessional velocity of the spin of a gyroscope in a
gravitational field is given by the general relation
(\cite{straumann} sec. 5.5)
\begin{equation}
\mathbf{\Omega =-}\frac{1}{2c^{2}}\left( \mathbf{v}\wedge
\mathbf{a}\right)
+G\frac{3\mathbf{r}\mathbf{r\cdot S}-\mathbf{S}r^{2}}{c^{2}r^{5}}+\frac{3}{2}%
G\frac{M}{c^{2}}\frac{\mathbf{r}\wedge \mathbf{v}}{r^{3}}
\label{girosc}
\end{equation}
The symbols $\mathbf{v}$ and $\mathbf{a}$ stand for the velocity
and (non-gravitational) acceleration three-vectors. All terms in
(\ref{girosc}) can, in principle, be measured even on Earth.
However, in the terrestrial environment, the first term, which is
the Thomas precession, dominates. In free fall, i.e. for an
orbiting gyroscope, there is no Thomas precession: only the second
and third term remain, representing respectively the
Lense-Thirring effect (depending on the angular momentum of the
Earth) and the geodesic or de Sitter precession. The latter, which
depends on parallel transport of a vector in curved space-time
\cite{desit} has a magnitude about 100 times greater than the
Lense-Thirring one. However for a gyroscope in a polar orbit, the
angular velocity of the Lense-Thirring precession stays always in
the revolution plane, while the angular velocity of the de Sitter
precession is perpendicular to that plane. It is expected that GPB
will measure both precessions.\newline Actually the mission will
carry four identical gyroscopes. Each gyroscope is a fused quartz
sphere coated with superconducting Niobium. The spheres float on a
magnetic field and are set and maintained into rotation by a
slight blow of Helium gas in an otherwise empty environment. The
precession of the axis of each sphere will be revealed by a SQUID
sensor. The expected value is 0.042 arcsec/year.

\emph{Fundamental implications of GPB}.
\addcontentsline{toc}{subsubsection}{Fundamental Implications of
GPB} Recently Nayer and Reynolds \cite{brane} pointed out that
Gravity Probe B could test some implications of the
Randall-Sundrum two-brane scenario: in fact this model predicts
that Earth originated gravitomagnetism should be some 16 orders of
magnitude smaller than the one predicted by general relativity,
hence GPB should see no gravitomagnetic effect at all. So, if a
non-zero precession is actually measured, the peculiar two-brane
scenario is somehow wrong. This is indeed an example of the
persistency of a misconception according to which gravitomagnetism
must still be verified. In fact, considering the other evidences
for the existence of gravitomagnetic effects of the GR predictions
size, the Randall-Sundrum two brane theory should already be
considered as being disproved. \newline

\emph{\normalsize The HYPER Project }%
\addcontentsline{toc}{subsubsection}{The HYPER Project } Thanks to
the big improvement in the accuracy of experimental devices, it is
possible to expect that in the future the effects of a rotating
massive body on quantum particles can be detected (for further
details on the quantum effects of rotation, see subsection
\ref{sec:coupling}) . In particular, the available cooling
techniques of gaseous atomic agglomerates makes it possible to
prepare interfering atomic beams, which stay for a long time
inside the interferometer and thus possess a long interaction
time: this technology may prove to be useful to detect the
Lense-Thirring effect too. A realization of this idea is attempted
within the HYPER project \cite{hyper}, which is planned to put
atomic interferometers in space with the aim of measuring the fine
structure constant, the quantum gravity foam-structure of space
time and the Lense-Thirring effect. This should be done using two
atomic interferometers based on Mg, and two based on Cs atomic
beams, placed in two orthogonal planes. The claimed resolution,
for an integration time of 1000 s, should be $10^{-14}\ $rad/s.
Contrary to GPB, where the cumulative effect over approximately
one year is read out, in the case of HYPER the gravitomagnetic
field would be measured locally, without integration over many
days, but only in a few minutes. HYPER is planned to fly within
the next 10 years.\newline

\subsection{Earth based experiments}

\label{ssec:earthbased}

\emph{\normalsize Effects on a Pendulum. }%
\addcontentsline{toc}{subsubsection}{Effects on a Pendulum}
Braginsky et al. \cite{brapotho} proposed to detect the
gravitomagnetic field of the Earth by studying its effect on the
plane of swing of a Foucault pendulum, collocated at the South
Pole. This experiment can be thought as an Earth based version of
the GPB mission. The conceptual bases are very simple: the
pendulum's mass $M$, swinging with velocity $\mathbf{v}$, is
subjected to the
gravitomagnetic field of the Earth at the pole $\mathbf{B}_{G}=4\frac{G}{c}%
\frac{\mathbf{S}}{R^{3}}$, which produces a force $\mathbf{F}_{GM}=m\frac{%
\mathbf{v}}{c}\wedge \mathbf{B}_{G}$. This fact causes a
precession of the principal axis of the pendulum, with respect to
a fixed azimuth, with angular velocity
\begin{equation}
\Omega _{GM}=\frac{B_{G}}{2c}=0.281\textit{''/yr} \label{eq:mech3}
\end{equation}
or, in terms of the angle of precession for a duration of the experiment of $%
\tau $:
\begin{equation}
\Delta \phi _{GM}=\Omega _{GM}\times \tau =0.046"\frac{\tau }{60d}
\label{eq:mech4}
\end{equation}
As the authors point out, the main sources of errors are:
frictional anisotropy, Pippard precession (due to variability of
the angular momentum of the pendulum with respect to the rotation
axis of the Earth, during the swing), frequency compensation,
antiseismic isolation for the pendulum, atmospheric refraction and
physical distortion for the telescope (used to point to a
reference star, with respect to which the precession is
evaluated), dynamic range for the readout system. Studies of these
sources have to be done accurately, in order to determine the
feasibility of the experiment, which could be a further
verification of the gravitomagnetic effect. \newline

\emph{Lense-Thirring in the lab.} %
\addcontentsline{toc}{subsubsection}{Lense-Thirring in the lab} A
ground-based test of the Lense-Thirring effect was proposed in
1988 by Cerdonio et al.\cite{cerdo}: they planned to compare the
astrometric measurements of the terrestrial rotation (as performed
by using the VLBI) with an inertial measurement of the angular
velocity of the laboratory. The latter would be obtained using a
new detector of local rotation, the so called Gyromagnetic
Electron Gyroscope. This GEG would be made of a ferromagnetic rod
rigidly inclosed in a superconducting shield, surrounded by a
SQUID sensor; the two parts of the apparatus would be differently
magnetized by rotation (Barnett and London effects) thus revealing
the sought for rotation of the laboratory \cite{cerdo1}. The
comparison should be performed off-line, which is one of the
advantages of their proposal. The experiment seems to be in the
range of\ feasibility, even though the problem of isolation from
seismic noise is serious (the possibility of using a spacecraft
has been considered).\newline

\section{Clock Effect and Gravitomagnetism}

\label{sec:clock}

\emph{The effect}. \addcontentsline{toc}{subsubsection}{The
effect} A gravitomagnetic field affects the motion of standard
clocks orbiting around a rotating mass and the proper time they
measure. For example, the angular velocity of a rotating test body
should be greater or lower depending on the direction of its
angular momentum, with respect to the angular momentum of the
source. Let us consider more closely the effect on the measure of
time, starting from the Kerr metric \cite{kerr}, in
Boyer-Lindquist coordinates \cite{boylin}:
\begin{eqnarray}
ds^{2} &=&(1-\frac{2\mu r}{\rho ^{2}})c^{2}dt^{2}-\frac{\rho ^{2}}{\Delta }%
dr^{2}-\rho ^{2}d\theta ^{2}-\left( r^{2}+a^{2}+\frac{2\mu
ra^{2}\sin
^{2}\theta }{\rho ^{2}}\right) \sin ^{2}\theta d\phi ^{2}  \nonumber \\
&&+\frac{4\mu ra\sin ^{2}\theta }{\rho ^{2}}d\phi cdt
\label{eq:kerr}
\end{eqnarray}

where $\mu =\frac{GM}{c^{2}}$, $\rho ^{2}=r^{2}+a^{2}\cos ^{2}\theta $, and $%
\Delta =r^{2}-2\mu r+a^{2}$, $a=S/Mc$. \newline Let us start from
a standard clock on a circular ($r=cost$) timelike
geodesic orbit in the equatorial plane. From the geodesic equation\emph%
{\normalsize \ }
\begin{equation}
\frac{d^{2}x^{\alpha }}{ds^{2}}+\Gamma _{\mu \beta }^{\alpha }\frac{dx^{\mu }%
}{ds}\frac{dx^{\beta }}{ds}=0  \label{eq:geo11}
\end{equation}
we obtain the only non trivial equation
\cite{mashh3}\emph{\normalsize \ }
\begin{equation}
(a^{2}-\frac{r^{3}}{\mu })d\phi ^{2}-2ad\phi cdt+c^{2}dt^{2}=0
\label{eq:CE1}
\end{equation}
Depending on the direction of rotation of the clock, i.e.
co-rotating or counter-rotating with respect to the rotation of
the source, we have two different proper revolution periods
measured by the clock after a closed orbit ($2\pi $ change in the
azimuthal $\phi $ angle). In terms of the proper time of the clock
one has:\emph{\normalsize \ }
\begin{equation}
\tau _{\pm }=T_{0}(1-\frac{3\mu }{r}\pm 2\frac{a}{c}\omega
_{0})^{1/2} \label{eq:CE2}
\end{equation}
where $T_{0}$ and $\omega _{0}$ are the Keplerian period and
frequency of the orbit, that is $\omega
_{0}=(\frac{GM}{r^{3}})^{1/2}$. So, the difference between the
squares of the proper times is\emph{\normalsize \ }
\begin{equation}
\tau _{+}^{2}-\tau _{-}^{2}=8\pi \frac{a}{c}T_{0}  \label{eq:CE3}
\end{equation}
This result is exact, i.e. approximation free; however, in most of
the interesting cases, it suffices to use the lowest
post-Newtonian order, so it turns out that\emph{\normalsize \ }
\begin{equation}
\tau _{+}-\tau _{-}\simeq 4\pi \frac{a}{c}  \label{eq:CE4}
\end{equation}

The result (\ref{eq:CE4}) is extremely appealing, because it is
independent from both the radius of the orbit and the
gravitational constant $G$. This result was found also by
Mitskevich and Pulido Garcia \cite{mit} and by one of us (A.T.)
\cite{tarta2},\cite{tarta5},\cite{tarta3}. Using a geometric
approach, based on the symmetry of space-time world lines of
objects rotating on circular trajectories, it is showed that, for
an axisymmetric metric tensor, the world lines are helices drawn
on the flat bidimensional surface of a cylinder, so they can be
easily studied from an euclidean point of view.\newline To give
some numerical estimates of the magnitude of this effect, we may
consider clocks orbiting around the Earth, and we obtain $\tau
_{+}-\tau _{-}\simeq 10^{-7}\ s$, which is well within the
capability of the clocks available at present. In the next
section, we shall analyze the difficulties which arise in
preparing actual experiments to detect this effect.\newline We
obtain a different result if we compare the proper times shown by
two counter-orbiting clocks at the second conjunction event along
equal radius trajectories in the equatorial plane
\cite{tarta2}:\emph{\normalsize \ }
\begin{equation}
\tau _{+}-\tau _{-}\simeq 8\pi \frac{\mu a}{cr}  \label{eq:CE6}
\end{equation}
Formula (\ref{eq:CE6}) gives, in the case of the Earth, a time
difference of
$\sim 10^{-16}\ s$, which is much smaller than the time gap we obtained in (%
\ref{eq:CE4}).\newline The proper time difference in clocks
orbiting a rotating massive body, can
be explained in terms of the gravitational Aharonov-Bohm effect \cite{mashh3}%
,\cite{tarta4}, where the phase of the radiation is influenced by
the presence of a vector gravitomagnetic potential.\newline For an
accurate analysis of the special classes of observers with regard
to
the gravitomagnetic clock effect see for example Bini et al. \cite{bini},%
\cite{bini1} and references therein.\newline

\emph{Orbiting
Clocks}.\addcontentsline{toc}{subsubsection}{Orbiting Clocks} The
use of orbiting clocks to detect the rotation effects in GR was
proposed by Cohen et al. \cite{corocl} in 1988, when the accuracy
and precision of atomic clocks was approaching the required
sensitivity, and the possibility to place such clocks on
satellites was in sight. They proposed to measure a
synchronization gap between clocks in co-rotating and
counter-rotating orbits, which they evaluated to be approximately
$1.92\times 10^{-17}\ $s (see section \ref{sec:sagnac}).\newline

\emph{\normalsize Gravity Probe C. }%
\addcontentsline{toc}{subsubsection}{Gravity Probe C} Rather than
considering the synchronization gap, Gronwald et al. \cite{gpc}
proposed to use clocks orbiting around the Earth to measure the
gravitomagnetic effect expressed by formula (\ref{eq:CE4}), which
is much larger than Cohen's gap,
and well within the measurement possibilities of modern atomic clocks \cite%
{lgm}. They studied a mission, called Gravity Probe C(lock), whose
aim is to detect directly the gravitomagnetic field of the Earth.
The name of this mission resembles closely to Gravity Probe B,
and, as we shall see in a
while, the fundamental nature of the effect they want to evidence is the same%
\footnote{%
For complete reference, we mention that the main result of the
Gravity Probe A mission was an accurate test of the gravitational
redshift, by means of
the sub-orbital flight of a rocket carrying a hydrogen maser clock \cite{gpa}%
,\cite{gpa1}.}.\newline In practice the time difference
(\ref{eq:CE4}) can be obtained by comparison of the proper times
shown by two clocks in co-rotating and counter-rotating orbits
around the Earth, after one sidereal revolution. The relative
gravitomagnetic variation of the orbital period with respect to
the Keplerian period $T_{0}$ is \emph{\normalsize \ }
\begin{equation}
\frac{\tau _{+}-\tau
_{-}}{T_{0}}=\frac{2S}{Mc^{2}}(\frac{GM}{r^{3}})^{1/2}
\label{eq:GPC1}
\end{equation}
which gives a numerical estimate of about $4\times 10^{-11}$ in
the case of
the Earth. Cohen and Mashhoon \cite{mashh3} observed that formula (\ref%
{eq:GPC1}) is the same as the relative gravitomagnetic precession
angle of Gravity Probe B gyroscopes, so it is expected that the
difficulties of the two experiments are equivalent. It is not
difficult to measure a time difference of $\simeq 10^{-7}\ $s with
modern technologies (future missions are planned to carry highly
accurate clocks in space \cite{clocks}, and the GPS can be used to
determine precisely the positions of the clocks), but as we are
going to show, the difficulties of the experiments originate
elsewhere.\newline First of all, the result (\ref{eq:CE4}) is
obtained considering two highly accurate and stable clocks, in two
accurate and stable orbits: in fact the (gravitoelectric)
post-newtonian corrections cancel exactly under the assumption
that the orbits are identical, and only the gravitomagnetic
correction remains \footnote{%
The effect for more general orbits, with small inclination with
respect to the equatorial plane is studied by Mashhoon \textit{et
al. }\cite{mashh5} and the problem of closeness of geodesic orbits
other than the equatorial ones is discussed. It is expected that
the effect decreases with increasing
inclination, and it becomes null for a polar orbit, because of symmetry.}.%
\newline
Other error sources can be summarized in two groups:\newline
\newline
i. errors due to the tracking of the actual orbits \newline
\emph{\normalsize \newline }ii. deviations from idealized orbits
due to \normalfont

\begin{itemize}
\item mass multipole moments of the Earth

\item radiation pressure

\item gravitational influence of the Moon, the Sun, and other planets

\item other systematic errors (such as atmospheric disturbances)
\end{itemize}

It seems that errors due to the tracking of the orbits are those
which require greater attention and careful study. In fact the
present accuracy in determining the position of the clocks in
space can be estimated (in terms of time difference) to be $\delta
t\simeq 10^{-6}\ $s \cite{gpc}, considering the available accuracy
in angle and position measurements: it is one order of magnitude
above the effect we want to measure. However it does not seem
impossible to achieve the required precision; moreover the clock
effect is cumulative so it can be made greater by performing many
orbits, thus overcoming the statistical tracking errors
too.\newline

\section{Correction to the Sagnac Effect}

\label{sec:sagnac}

When a light source and a receiver are located on a turntable,
which rotates with angular velocity $\omega $, the time for the
round trip of the light rays along a closed path varies with the
angular velocity itself. Moreover, the time will be different,
with fixed $\omega $, if the light beam is co-rotating or
counter-rotating. So, superimposing two oppositely rotating beams
leads to a phase difference and to interference, or in the case of
standing waves, to frequency shift and beats. This is the so
called Sagnac effect \cite{sagnac}, which can be explained in
terms of Special Relativity, on the basis of the break of
uniqueness of simultaneity in rotating systems
\cite{rizzi}.\newline
The time lag between the two light beams, up to the first order in $%
v^{2}/c^{2}$ is
\begin{equation}
\delta \tau _{S}=4\frac{A}{c^{2}}\omega  \label{eq:SA1}
\end{equation}
where $A$ is the area of the projection of the closed path
followed by the beams around the platform, orthogonal to the
rotation axis, and $\omega $ is the angular velocity of the
observer.\newline Today the Sagnac effect has many applications,
both for practical purposes and for fundamental physics; however
the improvement of the accuracy in measurements introduces the
need for precision corrections to the basic formula
(\ref{eq:SA1}), such as the general relativistic ones. The latter
were worked out by one of us (A.T) \cite{tarta}, considering the
case of the Kerr field, in the post newtonian approximation, and
so they can be interpreted as gravitomagnetic corrections,
originating from the off-diagonal term of the metric tensor. If
the ''rotating''\ platform is the Earth, and the circular path of
the light rays is around the equator, then the ''pure''\ Sagnac
delay corresponding to formula (\ref{eq:SA1}) is
\begin{equation}
\delta \tau _{S}=4.12\times 10^{-7}\ \mathit{s}  \label{eq:SA2}
\end{equation}
The correction depending on the pure mass term (Schwarzschild-like
correction) is
\begin{equation}
\delta \tau _{M}=4\pi \frac{\mu R}{c^{2}}\Omega \simeq 2.84\times
10^{-16}\ \mathit{s}  \label{eq:SA3}
\end{equation}
The first correction depending on the angular momentum of the
Earth is
\begin{equation}
\delta \tau _{a}=-8\pi \frac{a}{c}\frac{\mu }{R}\simeq -1.89\times
10^{-16}\ \mathit{s}  \label{eq:SA4}
\end{equation}
where the radius, the mass, the angular momentum per unit mass and
the angular velocity of the Earth were used; (\ref{eq:SA4}) in
fact coincides with (\ref{eq:CE6}). Other corrections are worked
out from the point of view of an observer on a geodetic orbit, but
the numerical estimates are not much different or better than the
ones we gave above. Anyway, it is reasonable that the fringe shift
produced by the general relativistic corrections is in
principle observable (with the time delays (\ref{eq:SA3}) and (\ref{eq:SA4}%
), using visible light, one obtains a $\sim 10^{-2}$ fringe
shift). Of course there are many technical difficulties in
performing such experiments, both on Earth and in orbit, but they
do not seem impossible to overcome in the next future.\newline

\emph{\normalsize Satellites ring. }%
\addcontentsline{toc}{subsubsection}{Satellites ring} An idea to
verify the GR corrections to the Sagnac effect is to make use of a
ring of orbiting satellites (such as those belonging to the GPS or
to the future European Galileo system). A stationary ring
configuration of satellites can be the way to force the light
beams to run in a closed circuit around the Earth, both in
co-rotating and counter-rotating direction. The time difference in
the propagation times should reproduce the effect expressed by formula (\ref%
{eq:CE6})-(\ref{eq:SA4}), once the much bigger classical Sagnac
effect has been subtracted out. Of course here too the technical
details need to be thoroughly worked out.

\section{Effects of the angular momentum on Michelson-Morley experiments}

\label{sec:MM}

Another manifestation of the effects of the angular momentum of
the source of gravity in interferometric measurements can be
studied considering the influence of the terrestrial rotation on
the famous Michelson-Morley experiment. This is a well known
experiment, which is often quoted in teaching the basic Special
Relativity. The idea of the experiment is the following: let us
imagine to place the interferometer on the equator, with one arm
along the South-North direction and the other arm along the
West-East direction. The light propagating in the interferometer
is subject to the influence of the gravitational field of the
Earth, which changes the propagation time with respect to the flat
space-time. Furthermore, the light propagating in the
East-West\emph{\ }arm should have a different time of propagation,
depending on whether it propagates in co-rotating or
counter-rotating sense.\newline Let us start from the null
space-time element, in an axisymmetric field
\begin{equation}
0=g_{tt}dt^{2}+2g_{t\phi }dtd\phi +g_{rr}dr^{2}+g_{\theta \theta
}d\theta ^{2}+g_{\phi \phi }d\phi ^{2}  \label{generale}
\end{equation}
We then suppose that the interferometer, placed near the equator,
has short enough arms.\emph{\newline }We can compute the
difference in propagation times between the two arms of the
interferometer, considering light world lines with $r=constant$
only. This choice corresponds to limiting the study to light beams
contained locally in a horizontal plane (this would actually
require a wave guide locally shaped as a constant gravitational
potential surface). It can be seen \cite{tarta6} that the
propagation time in the South-North-South arm is
\begin{equation}
t_{SNS}=t_{N}+t_{S}=2\sqrt{-\frac{g_{\theta \theta
}}{g_{tt}+2g_{t\phi }\Omega +g_{\phi \phi }\Omega ^{2}}}\Phi
\label{eq:SNS}
\end{equation}
while for the West-East-West arm we have
\begin{equation}
t_{WEW}=2\frac{\sqrt{g_{t\phi }^{2}-g_{tt}g_{\phi \phi
}}}{g_{tt}+2g_{t\phi }\Omega +\Omega ^{2}g_{\phi \phi }}\Phi
\label{eq:WEW}
\end{equation}
where $\Omega $ is the angular speed of the Earth and $\Phi =l/R$.
Here $l$ is the (proper) length of the arm of the interferometer
and $R$ is the radius of the Earth. We assume in general that
$l<<R$. Then, the difference in the time of flight along the two
arms at the lowest order of a weak field approximation is
\begin{equation}
\Delta t=t_{WEW}-t_{SNS}\simeq \frac{a^{2}}{R^{2}}\frac{l}{c}
\label{diff}
\end{equation}
The result (\ref{diff}) is obtained in the presumption that a
physical apparatus (bidimensional wave guide) obliges the light
rays to move along constant radius paths; the order of magnitude
estimate at the surface of the Earth for $1$ m long interferometer
arms is:
\begin{equation}
\Delta t\sim 10^{-20}\ \textit{s}  \label{numero}
\end{equation}
This effect is purely rotational and rather small but not entirely
negligible. \newline As we showed elsewhere \cite{tarta6} it is
possible to increase the value of (\ref{numero}), using
Fabry-Perot type interferometers, which permit multiple
reflections. Then the time difference between the two paths would
become $10^{-17}-10^{-16}$ s, which corresponds, for visible light, to $%
10^{-2}-10^{-1}$ fringe in the interference pattern.\newline The
obtained numeric value compares with the expected phase (and time)
shifts in the gravitational wave interferometric detectors now
under construction, as LIGO and VIRGO \cite{ligo1},\cite{virgo1}.
There indeed a
sensitivity is expected, in measuring displacements, of the order of $%
10^{-16}$ m which corresponds to a time of flight difference 4
orders of magnitude lower than (\ref{numero}) and consequently a
much higher sensitivity is required.\newline Of course the effect
as such is a static one, producing a DC signal and it would be
practically impossible to recognize its presence in the given
static interference pattern. On the other hand the spectacular
sensitivity of gravitational wave interferometric detectors is
obtained at frequencies in the range $10^{2}-10^{3}$ Hz. To
extract the information from the background and to profit of
highly refined interferometric techniques we would have to
modulate the signal. This result could be achieved, for instance,
steadily rotating the whole interferometer in the horizontal
plane. It is possible to design a configuration that, in
principle, should allow for the measuring of this effect, even if
a careful analysis of the technical details would of course be
needed in order to proceed further.

\section{Effects on signals propagation}

\label{sec:signals}

\emph{Shapiro delay and time asymmetry.}
\addcontentsline{toc}{subsubsection}{Shapiro delay and time
asymmetry} A well known effect on the propagation of light signals
in a gravitational field is the time delay: for example, a light
signal emitted from a source on the Earth toward another planet of
the solar system, and hence reflected back undergoes a time delay
during its trip (with respect to the propagation in flat
space-time), due to the influence of the gravitational field of
the Sun. Shapiro et al. \normalfont\cite{shapiro} measured this
time delay for radar-ecos from Mercury and Venus, using the
radio-telescopes of Arecibo and Haystack. Anderson et al.
\cite{ander} measured the time delay of the signals transmitted by
Mariner \normalfont6 and 7 orbiting around the Sun. Finally
Shapiro and Reasenberg obtained more accurate results using a
Viking mission that posed a transponder on the surface of Mars:
the theoretical prediction was verified within $\pm 0.1\%$
\cite{marte1} \cite{marte2}.

Now, we can consider here the correction to the time of
propagation due to the presence of a rotating source: the original
Shapiro measurements accounted just for the presence of a massive
source, described by the Schwartzschild solution, while we are
going to work out the time delay in post newtonian approximation
for a spinning source.\newline The metric tensor can be written in
cartesian coordinates, so that we have:
\begin{equation}
ds^{2}=g_{tt}dt^{2}+g_{xx}dx^{2}+g_{yy}dy^{2}+g_{zz}dz^{2}+2g_{xt}dxdt+2g_{yt}dydt
\label{eq:SH1}
\end{equation}
supposing the angular momentum of the source is parallel to the
$z$ axis. For a light ray propagating in the $xy$ plane we can
consider a first approximation, in which it propagates in a
straight line with $x=b=const$, from $y_{1}$ to $y_{2}$
($y_{1}<0$, $y_{2}>0$). Then (\ref{eq:SH1}) becomes
\begin{equation}
0=g_{tt}dt^{2}+g_{yy}dy^{2}+2g_{yt}dydt  \label{eq:SH2}
\end{equation}
It is easy to see that we can write, approximately, the coordinate
time of propagation from $y_{1}$ to $y_{2}$ in the form
\[
T(y_{1},y_{2})=T_{0}+T_{M}+T_{Ma}+...
\]
where $T_{M}$, $T_{Ma}$, are the corrections to the Minkowski
geometry propagation time $T_{0}$, and we neglect terms which are
smaller. In particular, $T_{M}$ is the delay effect measured by
Shapiro, while $T_{Ma}$ is the effect of the angular momentum we
want to outline. We obtain
\begin{eqnarray}
T_{0} &=&\frac{y_{2}-y_{1}}{c}  \label{eq:SH4} \\
T_{M} &=&\frac{2GM}{c^{3}}\ln \frac{y_{2}+\sqrt{b^{2}+y_{2}^{2}}}{y_{1}+%
\sqrt{b^{2}+y_{1}^{2}}}  \label{eq:SH5} \\
T_{Ma} &=&-\frac{2GMa}{c^{3}b}\left[ \frac{y_{2}}{\sqrt{b^{2}+y_{2}^{2}}}-%
\frac{y_{1}}{\sqrt{b^{2}+y_{1}^{2}}}\right]  \label{eq:SH6}
\end{eqnarray}
To give some numerical estimates, let us consider a realistic
situation in the solar system as, f.i., the propagation of signals
from the Earth to
Mercury; then $M=M_{\odot }$, $|y_{1}|=1.5\times 10^{11}\ $m, $%
|y_{2}|=6\times 10^{10}\ $m, $a_{\odot }\simeq 3\times 10^{3}\ $m, $%
b=6.8\times 10^{8}\ $m, hence
\begin{eqnarray*}
T_{0} &=&700\ \textit{s} \\
T_{M} &=&1.1\times 10^{-4}\ \textit{s} \\
T_{Ma} &=&1.0\times 10^{-10}\ \textit{s}
\end{eqnarray*}
The first angular momentum correction is six orders of magnitude
smaller than the main mass contribution, which has been measured
in the past. It is interesting to investigate whether this
correction could be measured in the context of a future space
mission, maybe exploiting the asymmetry of this effect. In fact
the sign of the correction changes depending on the fact that
light propagates in the $x>0$ or $x<0$ region. A possible way to
evidence this effect is proposed here, without entering into
details, which will be given elsewhere.\newline Let us imagine
light coming from far sources, which passes in the field of a
rotating object, a situation which is common in astrophysics.
Again, we are going to explain and evaluate the magnitude of this
effect, without giving the details, that will be found elsewhere.
\newline Let us consider the same metric tensor and coordinates of
the previous section, and two symmetric light rays propagating in
the $xy$ plane with constant $x$, that is $x^{+}=b$ and
$x^{-}=-b$. It is easy to show that, in the first post-newtonian
order, the time difference between the two signals, when received,
is
\begin{equation}
\delta \tau =-8\frac{GMa}{bc^{3}}  \label{eq:AS4}
\end{equation}
If we consider the Sun, or respectively Jupiter, as sources of the
gravitational field acting upon the light beams, and choose $b$ as
small as possible, that is of the order of the radius of the
source, we have the following numerical estimates:
\begin{eqnarray}
\delta \tau _{S} &=&1.6\times 10^{-10}\ \textit{s}  \label{eq:AS5} \\
\delta \tau _{J} &=&4.2\times 10^{-13}\ \textit{s}  \label{eq:AS6}
\end{eqnarray}
The time difference can be evidenced by superimposing the two
beams, and letting them interfere.

A good signal source could be a pulsar, because of its fine and
stable timing. It seems that the Sun is not that good
gravitational field source, because the light passing nearby it,
i.e. in its corona, is subject to many perturbations that may hide
the effect we want to see. We think that this
idea deserves some further investigation, because it seems somehow promising.%
\newline

Moreover, we want to point out that  Ciufolini \textit{et
al.}\cite{ciufolini02a},\cite{ciufolini02b} have recently studied
the problem of time delay due to spin, with particular attention
to the possibility of detecting the effect of
angular momentum by gravitational lensing.\\

In our calculation we have considered, for the sake of simplicity,
rectilinear propagation of light rays. We know, since the very
beginning of the GR tests, that this is not the case, because
light is bent by the gravitational field. It is interesting to
notice that the angular momentum of the source influences the
bending of light too. The magnitude of the bending angle $\delta
\phi $ turns out to be \cite{cobri}:
\begin{equation}
\delta \phi =\frac{4GM}{c^{2}b}\left( 1-\frac{1}{c}\frac{\mathbf{S\cdot n}}{%
Mb}\right)  \label{eq:AS7}
\end{equation}
where $b$ is the impact parameter, $\mathbf{S}$ is the angular
momentum of the source with mass $M$ and $\mathbf{n}$ is a unit
vector in the direction of the angular momentum of light about the
center of the source-body. The terms outside the parenthesis in
(\ref{eq:AS7}) is the pure Schwarzschild term, i.e. the
gravitoelectric contribution. The relative correction to the pure
mass term, in the case of the Sun, can be estimated in $10^{-6}$,
which is very small. This shows that our toy model, which
considers straight line propagation of light, is acceptable.

\section{Gravitomagnetic coupling between Spin and Angular momentum}

\label{sec:coupling}

The GEM framework allows us to extend some well known effects of
electrodynamics to the gravitational field. In fact, we have seen
that the gravitomagnetic field of the Earth is analogous to a
dipolar magnetic field. A spinning particle possesses a
''gravitomagnetic dipole moment'', i.e. a spin \mbox{\boldmath
$\sigma $}, which couples to the gravitomagnetic field of a
rotating mass with an interaction energy analogous to the magnetic
interaction $H=-\mathbf{m\cdot B}$, between a magnetic dipole
$\mathbf{m}$ and the magnetic field $\mathbf{B}$. The effects of
this coupling on the deflection of polarized radiation have been
studied in the past: it was found that the paths of right
circularly and left circulaly polarized photons, propagating in
the gravitational field of a rotating object, split because of the
coupling between the helicity and the angular momentum of the
source \cite{mashh7}\cite{mashh8}, much like in a Stern-Gerlach
experiment with polarized matter passing through an anisotropic
magnetic field. However the deflection angle (in the case of
photons of wavelength $\lambda $ propagating around the Sun with
distance of closest approach $b$)
\begin{equation}
\delta \phi \sim \lambda \frac{GS}{c^{3}b^{3}}  \label{eq:dphi}
\end{equation}
is too small and so it was not detected in the first experiments
done in the 70's \cite{har1}\cite{har2}.\newline The effect should
be observable for collapsed objects, thus being a way to measure
their angular momentum. In the case of neutrinos, gravity-spin
coupling should lead to a helicity flip, which should be important
for neutron stars and supernovae \cite{caipap}. Ahluwalia
\cite{ahlu} studied the influence of angular momentum-spin
coupling on quantum mechanical clocks, showing that they do not
always redshift identically when moved from the field of a non
rotating object to the field of a rotating source. The theoretical
investigations on the role of intrinsic spin in gravitational
interaction and the continuous improvement of precision
measurement techniques have enlarged the horizon of possibly
detectable effects. Let us
give orders of magnitude, following the treatment given by Mashhoon \cite%
{mashh2}, where more references and a more detailed discussion can be found.%
\newline
In analogy with the electrodynamic case, $H~=-\frac{\mathbf{\sigma }}{c}%
\cdot \mathbf{B}_{G}=~\mathbf{\sigma }~\cdot ~\mathbf{\Omega }$ is
the interaction Hamiltonian, where $\mathbf{\Omega }$ is the
angular precession frequency given in eq (\ref{eq:LT3}) and
\mbox{\boldmath $\sigma $} is the spin of the particle under
study. For an experiment performed on the surface of a rotating
body, just like the Earth, in a laboratory at latitude $\alpha $,
it turns out that the interaction energy difference between
particles of spin $\sigma =s\hbar $ polarized up and down (i.e.
perpendicular to the surface) is
\begin{equation}
E_{+}-\ E_{-}=4s\hbar \frac{GS}{c^{2}r^{3}}\sin \alpha
\label{eq:sc1}
\end{equation}
For the Earth and $s=1$, we have $\Delta E_{E}\simeq 2\times
10^{-29}eV$;
for Jupiter $\Delta E_{J}\simeq 10^{-27}eV$, and the same for the Sun $%
\Delta E_{S}\simeq 10^{-27}eV$. For a compact and fast rotating
body, such as a neutron star, we could have $\Delta E_{NS}\simeq
10^{-14}eV$.\newline Also light scattering around a rotating
object is influenced by the polarization of the incident
radiation, because a helicity-rotation coupling is present. It is
expected that right circularly polarized (RCP) and left circularly
polarized (LCP) waves be separated, in the gravitational
deflection, by the small gravitomagnetic splitting angle
(\ref{eq:dphi}). In addition to this effect, a rotation of the
plane of polarization along the ray is expected. Furthermore, due
to the different phase speed, the arrival times of positive and
negative helicity radiation originating near a
rotating object and propagating freely outward, are different\footnote{%
Of course, we do not know the absolute value of the propagation
times, because we do not know their origin; however it is possible
to measure, in principle, their difference.}:
\begin{equation}
T_{+}-T_{-}=-\frac{\lambda G\mathbf{S\cdot r}}{c^{4}\pi r^{3}}
\label{eq:sc3}
\end{equation}
where $\mathbf{r}$ is the distance between the emission point of
the
radiation and the center of the rotating source, whose angular momentum is $%
\mathbf{S}$; $\lambda $ is the wavelength of the
radiation.However, these
effects are very small, but it may have astrophysical implications \cite%
{ruffdam} and they may become interesting in microlensing with
polarized radiation \cite{micro},\cite{micro1}. \newline

The Hamiltonian $H={\mbox{\boldmath $\sigma $}\cdot
\mbox{\boldmath $\Omega $}}$ we introduced before, is position
dependent, so there is a
Stern-Gerlach force $\mathbf{F=-\nabla }H$ too, which acts on the particle%
\footnote{%
It is interesting noticing that the force on the particle is
mass-independent, hence the resulting acceleration is mass
dependent, thus violating the universality of gravitational
acceleration. Indeed, also the different deflection angle for RCP
and LCP radiation is a violation of the principle of equivalence
\cite{mashh10}. However these are wave effects, and the
universality of free fall is recovered in the JWKB limit.}. This
force causes a shift on the particle weight, which is different if
the spin is
polarized up or down. In fact, the effective weight for the particle is $%
W=mg-\mathbf{F\cdot \hat{r}}$, hence it depends on the
polarization:
\begin{equation}
W_{\pm }=mg\mp 6s\hbar \frac{GS}{c^{2}r^{4}}\sin \alpha
\label{eq:sc4}
\end{equation}
The ratio between the correction and the weight of a particle at
the surface of the Earth is $\sim 10^{-29}$. Again the effect is
too small to be measurable also in the next future.\newline

\section{Forces}

\label{sec:forces}

\emph{\normalsize Laboratory tests. }%
\addcontentsline{toc}{subsubsection}{Laboratory tests} In a paper
published in 1977, Braginsky et al. \cite{bct} proposed various
laboratory based experiments, to test relativistic gravity. Among
them, there were some tests of ''magnetic-type gravitational
forces'', i.e. gravitomagnetic effects, which originate from the
off-diagonal terms $g_{0i}$ of the metric tensor. They are
characterized by the sensitivity to the direction of rotation of
both the source and the detector. \newline One proposed experiment
is the gravitational analogue of the Amp\`{e}re experiment. The
source is an axially symmetric body, and the detector is a small
sphere, placed on the same axis as the main body. Both the source
and the sphere rotate about the common axis. According to the fact
that the rotation sense is the same or is opposite one has
repulsion or attraction. The detector mass is the end of an arm of
a torsion balance , which is used to measure the entity of the
force. For reasonable values of the parameters, the order of
magnitude of the force per unit mass ($m_{d}$) of the detector is
\begin{equation}
\frac{F}{m_{d}}\simeq 10^{-20}\textit{m/s}^{2}  \label{eq:mech1}
\end{equation}
The rotation is modulated at the eigenfrequency of the torque
balance.

Another laboratory experiment, is a variant of the Davies
experiment, which was proposed to measure the post-Newtonian
\textquotedblright dragging of the inertial
frames\textquotedblright\ caused by the rotation of the Sun:
two ideal light beams of infinitesimal wavelengths\footnote{%
Infinitesimal means that the wavelength is small compared to the
dimensions of the waveguide, i.e. in the geometric optics limit.}
travel in a thin toroidal waveguide, around the rotating
laboratory source. If the wave guide is kept at rest with respect
to a distant inertial frame, the standing wave pattern made by the
two waves should move relative to the guide with the angular
velocity $\Omega _{D}$, which represents the dragging of the
inertial frames. For a reasonable value of the mass of the source,
its size
and its velocity of rotation, one has $\Omega _{D}\simeq 6\times 10^{-21}$%
rad/s.

A time-changing gravitomagnetic field would produce a
gravitational electromotive force, and this fact can be used to
perform a Faraday-type experiment. The source of this field is a
cylinder, set into uniform rotation, and moved up and down along
its rotation axis. The detector consists in an axially symmetric
sapphire crystal, which is mounted coaxially with respect to the
source. The motion of the cylinder produces an induction
gravitoeletric field, which drives oscillations in the detector at
the same eigenfrequency of the source motion. The order of
magnitude of the driving force per unit mass $m_{d}$ is (for
reasonable values of the parameters)
\begin{equation}
\frac{F}{m_{d}}\simeq 10^{-20}\mathit{m/s}^{2}  \label{eq:mech2}
\end{equation}
or something less, according to the authors, since breaking
effects may occur. It is possible, however, that in the future, a
gravitational Faraday effect could be found in binary
pulsars.\newline

\emph{\normalsize Force on a Gyroscope. }%
\addcontentsline{toc}{subsubsection}{Force on a Gyroscope} In
Gravitoelectromagnetism it turns out that the force acting on a
gyroscope, in the field of a rotating object, can be calculated in
analogy with the expression of the force acting on a magnetic
dipole in a non-homogeneous magnetic field, and depends on the
direction of rotation. If the gyro has angular momentum
$\mathbf{L}$, the force acting on it, is \cite{ciufo}
\begin{equation}
\mathbf{F}=\frac{1}{2c}(\mathbf{L}\cdot \mathbf{\nabla
})\mathbf{B}_{G} \label{eq:Bforce}
\end{equation}
Out of the equatorial plane of the source a component of this
force will be radial. The order of magnitude is given by
\[
F\simeq \frac{G}{c^{2}}\frac{LS}{r^{4}}
\]
If we consider a prototype gyroscope on the Earth, with reasonable
mechanical parameters\footnote{%
For instance $l=1$\ m, $m=10\ $Kg, $\omega =10^{3}\ $Hz, where $l$
is the suspension length, $m$ is the gyroscope's mass, $\omega $
is its angular frequency.}, the order of magnitude of the force we
obtain is
\begin{equation}
F\simeq 10^{-17}\mathit{N}  \label{eq:Bforce3}
\end{equation}
This is a correction on the effective weight of the gyroscope,
and, in principle, it could be detected.\newline Indeed, we must
remember that in 1989, an anomalous weight reduction on a rotating
gyroscope was detected by Hayasaka and Takeuchi \cite{gw1}: a
decrease of the gyroscope's weight was detected only in its right
rotation, while the weight remained unchanged in its left
rotation, which ruled out the possibility of explaining it by
means of our GEM picture. It must be mentioned, however, that
afterwards different groups \cite{gw2},\cite{gw3}, performing the
same experiment, did not measure this mysterious reduction, which
remained unexplained.\newline

\section{Other effects}

\label{sec:other}

We list here a number of other ideas which have been proposed to
test gravitomagnetic effects.

\subsection{In space}

\label{ssec:inspace}

\emph{\normalsize Light propagation in the field of the Sun. }%
\addcontentsline{toc}{subsubsection}{Light propagation in the
field of the Sun} An approach similar to the one exploiting the
asymmetry in the Shapiro delay, which considers the propagation of
light signals in the gravitational field, can be found in the work
of Davies \cite{davi}. Here the possibility of measuring the
Angular Momentum of the Sun is studied, using the difference in
elapsed times for radio signals travelling clockwise and
counterclockwise around the Sun. These times are worked out for
free (geodesic) electromagnetic waves, and, depending on the range
of the solar rotation speed, the time difference is evaluated to
be between $1/6\times 10^{-10}\ $s and $1/6\times 10^{-8}\ $s,
which are not too different from the estimates we gave in
(\ref{eq:AS5}).\newline

\emph{\normalsize Tidal effects. }%
\addcontentsline{toc}{subsubsection}{Tidal effects} The
possibility of performing an orbital test of gravitomagnetism,
proposed by Braginski and Polnarev \cite{brapo1}, was critically
analyzed by Mashhoon and Theiss \cite{mashh6}: they studied the
relativistic corrections to the Newtonian tidal accelerations,
produced by a rotating system, and worked out the difficulties of
measuring these effects in orbit around the Earth, showing that
they are not different from those encountered by the Stanford's
Gravity Probe experiment.\newline Theiss \cite{thei} proposed
another test of gravitomagnetism: he considered the possibility of
measuring the effect of the terrestrial rotation on the tidal
acceleration between test masses in a satellite. He showed that
this effect should be unexpectedly large, and well within the
possibility of detection by a satellite experiment proposed by
Paik \cite{pai}. The aim of the mission was to measure, to very
high accuracy, the gravitational (i.e. gravitoelectric) gradient
of the Earth; the theoretical principles of detection of the GEM
field by means of a superconducting gradiometer were worked out by
Mashhoon et al. \cite{pai1}.\emph{\normalsize \newline }

\emph{\normalsize Effects on the orbits. }%
\addcontentsline{toc}{subsubsection}{Effects on the orbits} De Felice \cite%
{defeli} remarked that in the gravitational field of a spherical
source General Relativity implies the presence of a radial thrust
other than the centrifugal force, appearing when the angular
velocity of a non-geodesic circular trajectory is changed. In
order to maintain the same trajectory a correction would be
needed; measuring this correction would be a means of detecting
the GR effects. The required precision in the determination of the
parameters of an experiment in the Terrestrial environment would be $%
10^{-11} $ for the angular velocity and $10^{-13}$ in the value of
the mass.

\subsection{On Earth}

\label{ssec:onearth}

\emph{Effects on superconductors}.
\addcontentsline{toc}{subsubsection}{Effects on a superconductor}
The effects of the gravitomagnetic field on pure superconductors
were studied, by Li and Torr \cite{litorr}, with particular
interest on the interplay between the magnetic and the gravitomagnetic field.%
\newline
The conclusion was that in a pure superconductor always exists a
residual magnetic field produced by the gravitomagnetic field of
the Earth. This residual magnetic field produces in turn a local
perturbation of the gravitomagnetic external field.\\

\emph{\normalsize Effect on a spin $1/2$ system. }%
\addcontentsline{toc}{subsubsection}{Effect on a spin $1/2$
system} Another approach to the role of the gravitomagnetic field
of the Earth on a quantum system with spin $1/2$ is found in
\cite{cmcho}, where the Rabi formula for two-level system
transition is obtained, and the possibility of obtaining the
quantum Zeno effect is investigated.\newline

\section{Conclusions and the future}

\label{sec:conclusion}

\emph{\normalsize \ }As we have seen, in approximately 85 years of
research a number of proposals have been put forward concerning
the possibility to reveal the effects of gravitomagnetism. For a
long time many of these ideas have kept an essencially speculative
character, because in general the sought for effects are extremely
weak within the solar system. Strong effects in the vicinity of
neutron stars or black holes suffer of the fact that in systems
allegedly composed of those objects a variety of other effects are
present too and many parameters must be taken into account, so
that it is not easy to draw uncontroversial conclusions.

Nowadays however the improvement in length and time measurement
techniques, both in the lab and in space, have pushed the border
of detectability somewhat below the size of part of the effects
listed in this paper. Furthermore a number of space missions are
under way or being planned, with different purposes, which could
be fit for experiments on gravitomagnetism. In fact the cost of a
dedicated mission is quite high, however in many cases it is
possible to use the data collected from satellites sent for
different purposes. This is in general the case for any spacecraft
that carries transmitters, receivers and clocks. In the near
future the European Galileo project will be realized. The
satellites of those project, being designed for precision timing
and positioning of the satellites themselves or of ground based
stations, may easily be considered as ideal to reveal anisotropies
in the propagation of electromagnetic signals in the terrestrial
environment.

\end{document}